# Towards improved measurements of parity violation in atomic ytterbium

D. Antypas[1] • A. Fabricant[2] • L. Bougas[2] • K. Tsigutkin[3] • D. Budker[1,2,4]

**Abstract**

We report on progress towards performing precision measurements of parity violation in Yb, in which the theoretical prediction for a strong weak-interaction-induced effect in the $6s^2\ ^1S_0 \rightarrow 5d6s\ ^3D_1$ optical transition at 408 nm has already been confirmed, with a measurement of the effect at the ≈10 % level of accuracy. With a new atomic-beam apparatus offering enhanced sensitivity, we are aiming at precisely determining the parity violation observable in Yb, which will allow us to probe the distributions of neutrons in different isotopes, investigate physics beyond the Standard Model, as well as to study intra-nucleus weak interactions, through an observation of the anapole moment of Yb nuclei with nonzero spin.

We present the experimental principle employed to probe atomic parity violation, describe our new apparatus, and discuss the attained experimental sensitivity as well as the methods for characterizing systematics in these measurements.

## 1. Introduction

The Yb parity violation (PV) experiment at the University of California, Berkeley succeeded in 2009 in measuring the largest parity violation(PV) effect ever observed in any atomic system [1,2], about 100 times larger than that of Cs, in which the most precise measurement to date has been performed [3,4]. Responsible for the size of the effect is a large weak-interaction-induced mixing of two opposite-parity states, the 5d6s $^3D_1$ and 6s6p $^1P_1$ excited states of Yb (Fig. 1). Small energy spacing between these and favorable configuration mixing in the 6s6p $^1P_1$ state result in strong mixing of the $^3D_1$ and $^1P_1$ states, as noted in [5,6], where a prediction for the size of the effect was made.

Although this first observation of the Yb PV-effect at Berkeley confirmed the theoretical prediction of [5,6], the attained accuracy in the PV-related observable was at the 10% level. With a new apparatus recently constructed at the Helmholtz-Institut Mainz in Germany, we have enhanced the experimental sensitivity to a degree that will enable us to make precision measurements (sub-1%) of the Yb-PV effect and probe low-energy nuclear physics and potentially physics beyond the Standard Model. At the same time, another PV experiment, in Dy, is being restarted in Mainz [7].

Yb is a good platform for measuring atomic PV, since in addition to its large PV-effect, there are several naturally abundant isotopes ($^{168}$Yb-0.13%, $^{170}$Yb-3.02%, $^{171}$Yb-14.22%, $^{172}$Yb-21.75%, $^{173}$Yb-16.1%, $^{174}$Yb-31.9%, $^{176}$Yb-12.89%). Assuming that the dependence of the weak charge of the nucleus on the number of neutrons is as predicted by the Standard Model [8], a measurement of the variation of the PV-effect along an isotopic chain would provide information about the variation of neutron distributions among the Yb nuclei [9] and would aid nuclear theory in this sector. Conversely, isotopic-ratio measurements, combined with recent advances in calculations of neutron distributions, would enable probing physics beyond the Standard Model at the tree level with enhanced sensitivity [10]. Measuring the difference of the PV-effect among the hyperfine components of the same nonzero-spin isotope ($^{171}$Yb, nuclear spin $I$=1/2 and $^{173}$Yb, $I$=5/2), would yield information about the anapole moment of the Yb nuclei, and shed light on the parameter space of weak meson-nucleon couplings [8, 11], which within the current framework of the intra-nucleus weak interaction, are poorly understood. Observation of the nuclear-spin-dependent PV-effect may also reveal an additional spin-dependent contribution related to the nuclear quadrupole, as it was recently pointed out in [12].

Precise determinations of Yb spin-independent PV-effects could additionally serve as a probe of dark matter. A deviation of the Weinberg angle from the Standard Model prediction could be explained by the existence of a light dark-boson [13]. Finally, detection of an oscillating PV-effect would be a probe of cosmic parity violation arising from PV-coupling of a dark-matter field to matter, as it has been proposed in [14].

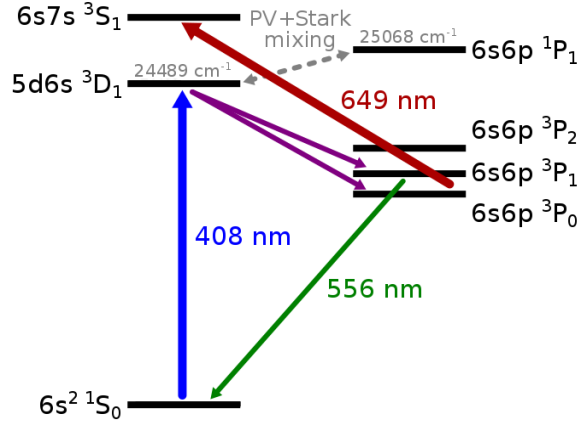

**Fig. 1** Yb energy levels and transitions related to the PV experiment.

## 2. Principle of the Yb experiment

The goal of the experiment is to measure the amplitude of the small PV-induced electric dipole transition between two states of nominally same parity: the $6s^2\ ^1S_0$ ground- and the $5d6s\ ^3D_1$ excited state of Yb, while resonantly exciting this optical transition in an atomic beam. This transition becomes slightly allowed since the excited state, due to PV-mixing with the $6s6p\ ^1P_1$ state, does not have a well-defined parity. As the PV-amplitude is too small for a direct detection of the corresponding transition rate, amplification of the experimental signal is performed using an interference technique [15]: a static electric field is applied to the atoms, inducing Stark-mixing between the $^3D_1$ and $^1P_1$ states, and resulting in a Stark-induced transition amplitude which with a properly chosen experimental field geometry, is allowed to interfere with the much weaker PV-amplitude. The resulting excitation rate contains a term due to this interference that can be extracted from the background of the much larger Stark-rate by performing electric field reversals. More reversals are possible, and are employed for systematics checks.

The Stark-PV interference rate is a spatial-inversion-odd, time-reversal-even pseudo-scalar quantity, which for the field geometry employed in our experiment (shown in Fig. 2), has the form:

$$(\boldsymbol{\varepsilon} \cdot \mathbf{B})[(\mathbf{E} \times \boldsymbol{\varepsilon}) \cdot \mathbf{B}]. \qquad (1)$$

Here $\boldsymbol{\varepsilon}$ is the optical field, $\mathbf{B}$ is a dc magnetic field and $\mathbf{E}$ is the static field. In this field configuration, the amplitude of a magnetic dipole transition (M1), which although suppressed, is still much larger than the PNC amplitude [16], is made to be out-of-phase with the dominant Stark amplitude, so that the Stark-M1 interference only occurs to the extent that field imperfections are present in the experiment. Employing a standing-wave optical field for the forbidden transition excitation provides additional suppression, since the M1-Stark interference term is of opposite sign for the two counter propagating components of light, so that the overall amplitude vanishes.

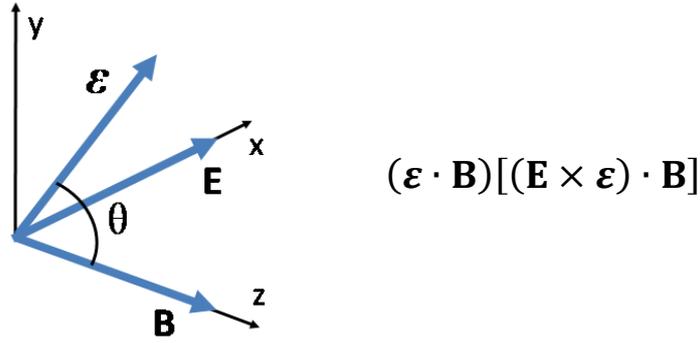

**Fig. 2** Experimental field geometry used in the Yb PV experiment and corresponding rotational invariant for the Stark-PV interference. The magnetic field **B** is applied collinearly with the axis of the atomic beam and the optical field of amplitude $\boldsymbol{\varepsilon}$ propagates collinearly with the electric field **E**.

The amplitude of a Stark-induced transition between two states with $J=0$ and $J'=1$, having total angular momentum $F$ and $F'$ with projections $m$ and $m'$ is of the form [15]:

$$A^{St}_{FmF'm'} = i\beta_{FF'}(-1)^{-q}(\mathbf{E} \times \boldsymbol{\varepsilon})_{-q}\langle F, m, 1, q | F', m' \rangle, \qquad (2)$$

while the PV-induced amplitude between the same states is given by:

$$A^{PV}_{FmF'm'} = i\,\zeta_{FF'}(-1)^{-q}\boldsymbol{\varepsilon}_{-q}\langle F, m, 1, q | F', m' \rangle. \qquad (3)$$

In (2) and (3), $\beta_{FF'}$ is the vector polarizability of the transition, $\zeta_{FF'}$ characterizes the PV-induced transition amplitude, $q=m'-m$ is the spherical component, and $\langle F, m, 1, q | F', m' \rangle$ a Clebsch-Gordan coefficient.

We now proceed with discussing the expected signatures of the PV-effect in the excitation rate induced by 408 nm light. We do this for the case of nuclear spin-zero ($I=0$) Yb isotopes. Analysis for non-zero spin isotopes has been carried out in [17], and proceeds along similar lines. There are three components of the transition ($m=0 \rightarrow m' = -1, 0, 1$), which are Zeeman-split upon the application of a magnetic field. For a field large enough to resolve these components, the transition rate for the $m$-component can be expressed as:

$$R_m \sim \left| A_m^{St} + A_m^{PV} \right|^2, \tag{4}$$

or explicitly for the individual Zeeman levels:

$$R_0 = 2\beta^2 E^2 \sin^2\theta + 4\zeta\beta E \sin\theta\cos\theta, \tag{5}$$
$$R_{\pm 1} = \beta^2 E^2 \cos^2\theta - 2\zeta\beta E \sin\theta\cos\theta, \tag{6}$$

where $\theta$ is defined in Fig. 2. We have omitted in the above expressions the extremely small PV-induced rate proportional to $\zeta^2$. The sign of the Stark-PV interference term (attaining the largest magnitude at a polarization angle $\theta=\pm\pi/4$) reverses sign upon reversing **E** or a rotation of $\theta$ by $\pm\pi/2$, allowing extraction of the PV-related signal from overall excitation rate. The interference term is of opposite sign for the $m=0$ and the $m=\pm 1$ components of the transition. The Berkeley experiment made measurements on all three Zeeman lines; in our experiment in Mainz however, we are planning to precisely measure the PV-effect on the $m=0 \rightarrow m'=0$ component only. This approach reduces the sensitivity to systematic effects in the particular component, compared to the other two.

As a measurement of the 408 nm transition rate by fluorescence detection from the interaction region suffers from poor geometrical collection efficiency, we use a shelving technique to enhance efficiency. Atoms in the Yb beam that underwent the forbidden transition decay with a 65% probability to the 6s6p $^3P_0$ meta-stable state. Downstream from the primary interaction region, these atoms are further excited to the 6s7s $^3S_1$ state using light at 649 nm. Fluorescence from the decay back to the 6s6p $^3P_0$, 6s6p $^3P_1$ and 6s6p $^3P_2$ states is detected, as well as from atoms relaxing to the ground state from the 6s6p $^3P_1$ state with a 556 nm photon emission.

An average of two photons are registered per atom initially decaying to the meta-stable state. The overall detection efficiency is of order 1/2 for the 408 nm transitions.

The Stark-PV interference term is extracted from the overall 408 nm transition rate by sinusoidal modulation of the electric field applied to the atoms. The resulting rate can be expressed as the sum of three terms: the Stark-PV term oscillating at the modulation frequency $\omega$, and two terms due to the dominant Stark rate: a dc term and a term oscillating at the 2$^{nd}$ harmonic $2\omega$. For the $m=0 \rightarrow m'=0$ transitions, these terms are as follows:

$$R^0 = \beta^2 E_0^2 \sin^2\theta, \tag{7}$$
$$R^{1st} = 4\zeta\beta E_0 \sin\theta\cos\theta\cos(\omega t), \tag{8}$$
$$R^{2nd} = \beta^2 E_0^2 \sin^2\theta \cos(2\omega t). \tag{9}$$

$E_0$ is the amplitude of the electric field. Phase-sensitive detection of the transition rate at both frequencies ($2\omega\ and\ \omega$) is employed using lockin amplifiers. Taking the ratio of amplitudes of the 1$^{st}$ to 2$^{nd}$ harmonic of the excitation rate, yields the PV-amplitude $\zeta$, scaled to $\beta E_0$:

$$r = \frac{R^{1st}}{R^{2nd}} = \frac{4\zeta}{\beta E_0} \cot\theta. \tag{10}$$

The advantage of normalizing the Stark-PV interference rate to the Stark-rate is that, parameters such as atomic-beam density and optical intensity need not be calibrated, and their fluctuations during the course of a measurement cancel out in the ratio. For a typical electric field of amplitude $E_0$=2 kV/cm applied to the atoms, $\theta=\pm\pi/4$, and using the measured value $\zeta$=39 mV/cm [1,2], we find for the expected value for the ratio: r= 7.8·10$^{-5}$, or 78 parts per million (ppm).

### 3. Apparatus

Following the relocation of the Yb experiment from UC, Berkeley to Mainz, a new atomic beam apparatus was constructed, aiming at improved measurements of the PV-effect. In essence, the apparatus consists of two-stage vacuum chamber: a small (oven) chamber that houses an oven that produces an effusive atomic-beam, and a larger (interaction) chamber, where the interaction of Yb atoms with the 408 nm light occurs in the presence of well-defined electric and magnetic fields, as well as the detection of the 408 nm excitations. We show a schematic of the atomic beam setup in Fig. 3, and in Fig. 4 a simplified schematic of the complete apparatus.

The Yb oven consists of a hollow stainless-steel cylinder, fitted with a nozzle made of an array of hypodermic tubes. The oven is heated to 550 °C in its rear section, while the front is maintained at °630 C. This temperature gradient is imposed to prevent nozzle clogging. The oven chamber is pumped with an 80 L/s turbo pump, and with the oven heaters running, reaches a base pressure of $3 \cdot 10^{-6}$ mbar.

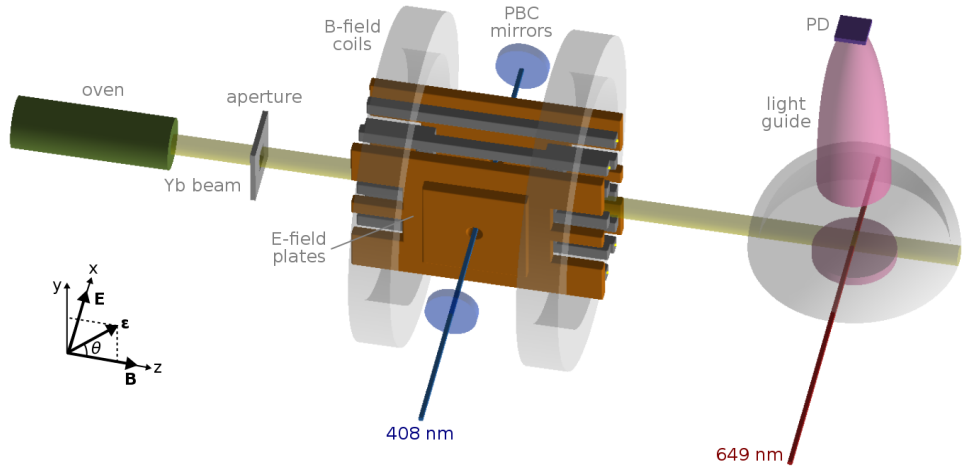

**Fig. 3** The Yb atomic-beam apparatus

Atoms effusing from the oven travel approximately 15 cm before passing through a rectangular 10 mm wide and 3mm tall aperture and entering the main (interaction) chamber. This aperture defines the shape of the atomic-beam in the interaction region and blocks atoms with large transverse velocities. The chamber is fabricated out of aluminum, as part of the effort to reduce stray magnetic fields in the experiment. It is a box of inner dimensions $45 \times 40 \times 30$ cm$^3$, pumped with a 240 L/s turbo pump. The residual pressure in this chamber is $1.5 \cdot 10^{-6}$ mbar.

After entering the main chamber, atoms travel another 12 cm before reaching the interaction region, with a mean longitudinal speed of 290 m/s and a transverse velocity spread (Full Width at Half Maximum-FWHM) of 8 m/s. In this region atoms intersect at normal incidence with the standing-wave light of an in-vacuum Fabry-Perot resonator, used primarily as a power-build-up cavity (PBC), but also to suppress the effects of the M1 transition. The PBC consists of a pair of 1 m radius of curvature (r.o.c.) mirrors, mounted on precision mounts and spaced by 28 cm, resulting in a 536 MHz free spectral range (FSR) for the cavity and a $1/e^2$ intensity waist of 212 μm. One of the mirrors is mounted on a piezo-electric transducer (PZT) that allows for precise control of cavity length. The two mirror mounts are attached to a quartz rod fitted to a heavy (20 kg) lead base that rests on 4 sorbothane pads, ensuring good vibration isolation of the system. The mirrors were purchased from Advanced Thin Films and have a reflectivity of 99.5% and losses <0.1%. The finesse of the PBC was measured to be ≈500. Since at the high intra-cavity power level required for the PV experiments there is a distortion in the lineshape of the 408 nm transition (occurring due to the effects of an off-resonant ac-Stark shift, and studied extensively in [18,19]), the choice of r.o.c. for the PBC mirrors was a compromise between the need for a cavity waist which is as large as possible and ease of alignment for the light coupled to the $TEM_{00}$ mode of the PBC. With proper mode-matching to the $TEM_{00}$, more than 100 W of near-UV light can be circulated in the PBC. Nevertheless, since the 408 nm transition lineshape broadens faster than it increases in amplitude at these high intensity levels, the cavity is typically operated with about 30 W of circulating power, corresponding to an interaction region average intra-cavity intensity of 21 kW/cm$^2$.

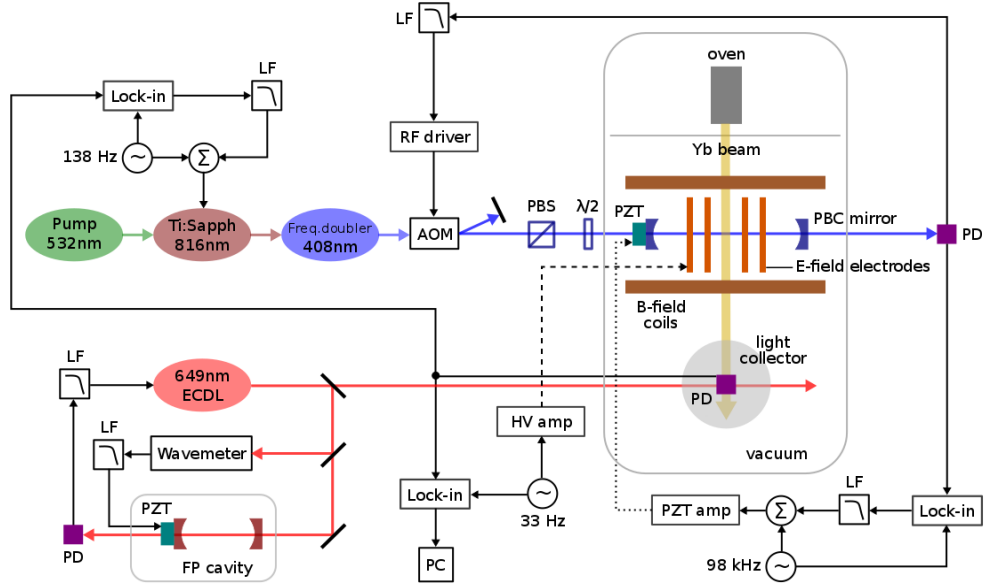

**Fig. 4** Simplified schematic of the complete setup of the Yb experiment. LF: loop filter, PBS: polarizing beam splitter, PD: photodiode, FP: Fabry-Perot cavity, PZT: piezo-electric transducer, λ/2: half-wave-plate, PBC: power build-up cavity, ECDL: external cavity diode laser.

Atoms in the interaction region experience the electric field created with a set of parallel field plates. The typical field amplitude used is 2kV/cm at a frequency of 33 Hz, with nonuniformity of less than 0.1% within the 5×5×5 mm$^3$ interaction region. These plates are 10×10 cm$^2$ in size, spaced by 5.505(2) cm and coated with gold. A set of 8 gold-coated electrodes with the shape of a rectangular bar is placed in the periphery of the main plates and used to produce transverse fields in either dimension perpendicular to the main field, with a nonuniformity ~ 5% within interaction volume. In addition, these electrodes serve to increase the primary electric field uniformity as well as to shield the interaction region from surfaces adjacent to the field plates. Transverse fields are not applied during the actual PV measurements, but are employed for systematics studies. The main field plates are biased symmetrically with the outputs of two high-voltage (HV) amplifiers (TREK 609B). The auxiliary field electrodes are biased with a system of precision voltage dividers to which voltage from the HV-amplifiers is applied. A set of an additional four HV-amplifiers (Ultravolt 4HVA) is used to add to the bias of these electrodes when either dc- or ac-transverse fields need to be applied in the interaction region.

The dc magnetic field required in the experiment is produced with a set of in-vacuum coils, having roughly Helmholtz geometry. The coils are made of round copper frames onto which copper wire is wound; the windings have an average diameter of ~19 cm and are spaced by 10 cm. The coil frames are water-cooled to prevent heating of the assembly under vacuum. The typical field applied to the atoms (collinear with the atomic-beam propagation direction) is 100 G. Several sets of coils external to the vacuum chamber are used to null the residual dc field in the interaction region, as well as to apply auxiliary fields for systematics checks.

Downstream from the interaction region, atoms enter the detection region, where they intersect with 649 nm light, used to detect the 408 nm excitations, as described in Section 2. About 2 mW of light are used to excite the atoms, with a corresponding intensity of ~7 mW/cm$^2$, high enough to saturate the 649 nm transition. The resulting fluorescence is collected with an in-vacuum light collector, guided out of the vacuum with a plexiglass light-pipe, and directed to the surface of a large-area photodiode (10×10$^2$ mm active area), whose current is detected using a 1 GΩ transimpedance amplifier which has a 700 Hz bandwidth.

The light coupled to the PBC is produced with a commercial system consisting of a frequency-doubled Ti:Sapphire laser (M-Squared SolStiS+ECD-X), pumped with a 15 W 532 nm laser (Lighthouse

Photonics Sprout G15). This system outputs more than 1 W of continuous wave (cw) light at 408 nm and has a specified linewidth < 100 kHz for the near-UV light. We find that when it is used to excite the ~ 28 MHz wide 408 nm resonance, the actual laser frequency noise adds an insignificant contribution to overall noise level in the detection of the small Stark-PV rate. Since the laser is well frequency stabilized, we use it as a short-term frequency reference. In order to keep the PBC resonant with the 408 nm laser, we lock the former to the latter. This is done by frequency-modulation spectroscopy: the cavity PBC length (and the corresponding resonance frequency) is modulated by dithering the cavity PZT, and an error signal is produced by demodulating the PBC transmitted signal with a lockin amplifier. The error signal is fed through an integrator back to the PZT. The bandwidth of this feedback loop is limited to a few kHz, on account of a PZT resonance at ~3 kHz.

In order to reduce residual PBC power fluctuations to an acceptable level, the intra-cavity power needs to be stabilized. This is done by measuring with a photodiode the PBC transmitted power and feeding it to a commercial PID controller, which controls the level of RF-power driving an acousto-optic modulator placed at the output of the laser system. With this scheme, we are able to reduce the PBC residual noise to a level completely insignificant compared to the shot-noise of the forbidden transition.

Although the 408 nm laser system provides a good short-term frequency reference, its long-term drifts (20 MHz/hr) are unacceptably large for acquiring data. We use the 408 nm resonance as an absolute long-term frequency reference to which the laser is stabilized. To achieve this, the frequency of the reference cavity of the Ti:Sapphire laser (to which the laser is stabilized), is modulated at 138 Hz, slow enough so that the PBC lock to the laser can follow. This modulation appears as a modulation in the 408 nm transition rate, and therefore in the fluorescence signal collected from the detection region. By demodulating this signal with a lockin amplifier, an error signal between the 408 nm resonance and the laser frequency is derived. This is then fed through a slow integrator back to the Ti:Sapphire laser reference cavity with a bandwidth of ~ 0.5 Hz, enough to maintain a stable lock of the laser to the peak of the transition.

The 649 nm light is produced with a commercial external cavity diode (ECDL) laser (Vitawave ECDL-6515R), outputting 10 mW of cw light. The laser is frequency-locked to the side-of-fringe of a home-built confocal Fabry-Perot cavity, in order to reduce fast frequency fluctuations. To ensure long-term stability, the laser frequency is read with a wavemeter (High

Finesse WSU2), and computer-based PID control is used to control the FP-cavity length, so that the laser frequency remains within 0.5 MHz of peak of the 70 MHz wide 649 nm transition. This two-stage stabilization scheme ensures that the 649 nm laser is both short- and long-term frequency stable. The contribution of the ECDL frequency noise to the overall noise in the PV detection is negligible.

The measurement of the amplitudes of the 1$^{st}$ and 2$^{nd}$ harmonics present in the 408 nm excitation rate is performed by demodulating the collected fluorescence signal from the detection region with a quad-phase lockin amplifier (Signal Recovery 7265), capable of simultaneous detection of two harmonic frequencies. Fig. 5 shows a spectrum of the 408 nm transition, recorded by measuring the 2$^{nd}$ harmonic of the transition rate with the lockin, while scanning the Ti:Sapphire laser frequency.

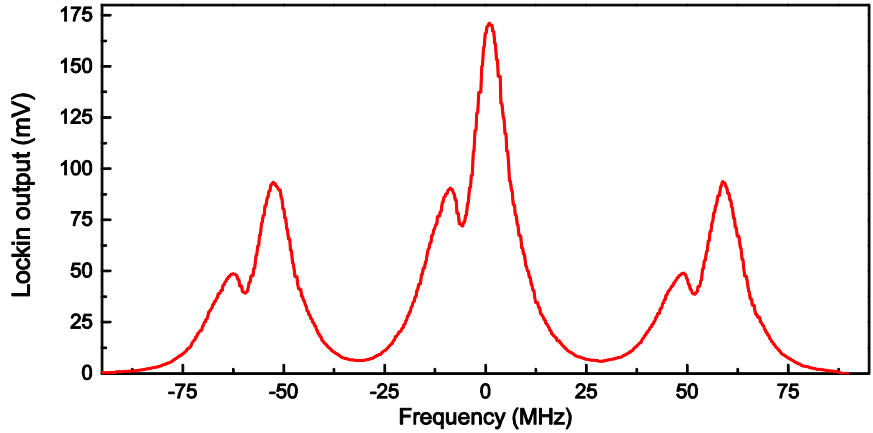

**Fig. 5** Amplitude of the 2$^{nd}$ harmonic of the $^{174}$Yb 408 nm transition rate (Stark-rate), measured with the lockin amplifier (time constant=50 ms), while scanning the Ti:Sapphire laser frequency. The 80 G magnetic field applied to the atoms splits the transition into its three Zeeman components. The observed lineshape asymmetry arises in the presence of the intense standing-wave field of the PBC, due to the off-resonance ac-Stark shift [18,19].

## 4. Experimental sensitivity

Sensitive measurements of the small PV-effect present in the 408 nm excitation rate require a signal-to-noise ratio (SNR) in its detection which is as large as reasonably possible. Aside from obtaining good statistics in the Stark-PV rate detection, this is particularly important for studies of systematics, since these are expected to require a substantial amount of data acquisition time. The attained 408 nm excitation rate in the new apparatus is about five times larger than that of the Berkeley setup; whereas the obtained SNR in its detection is improved by a factor of 30.

The apparatus measures the ratio of 1$^{st}$ to 2$^{nd}$ harmonics of modulation in the 408 nm excitation rate (Eq. 10). For the typical experimental conditions (ac-electric field of amplitude 2 kV/cm at $f$=33 Hz, polarization angle $\theta=\pm\pi/4$) the ratio corresponds to a ~ 18 µV rms level for the 1$^{st}$ harmonic (Stark-PV rate) at 33 Hz, over a ~ 250mV rms rate of the 2$^{nd}$ harmonic (Stark-rate) at 66 Hz. From noise measurements on the excitation rate for different ac-field amplitudes, we find approximately equal contributions to the overall noise level from shot- and technical noise (i.e. noise proportional to the signal), while the total background contribution is a few times smaller. We show in Fig. 6 a plot of the harmonics ratio, recorded over a period of 19 s, with the laser frequency stabilized to the peak of the $^{174}$Yb $m=0 \rightarrow m'=0$ component of the 408 nm transition. Based on the observed statistical fluctuations, we deduce an SNR of ~1 for the expected PV effect (~80 ppm) in one second of integration time.

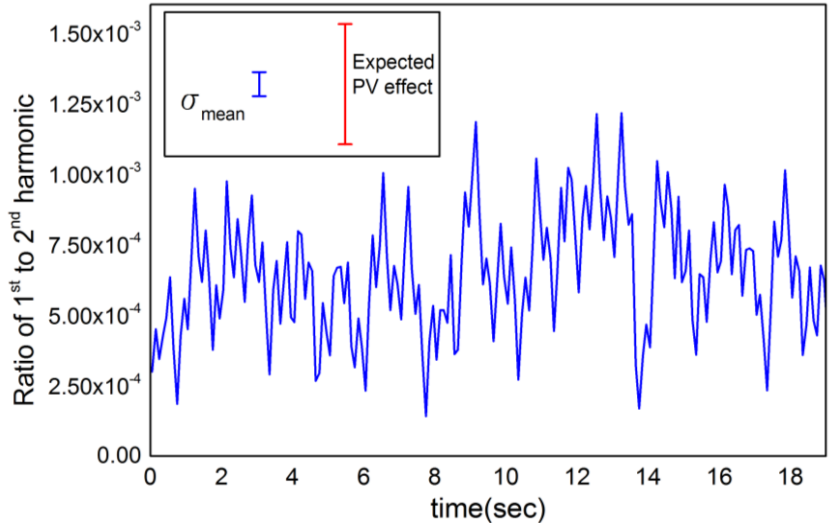

**Fig. 6** Ratio of the 1$^{st}$ to 2$^{nd}$ harmonic of modulation in the rate of the $m=0 \rightarrow m'=0$ component of the $^{174}$Yb 408 nm transition, measured over a period of 19 s with the laser stabilized to the peak of the transition. The 1$\sigma$ error in the mean value of the recorded ratio is ~ 14 ppm, corresponding to an uncertainty of 70 ppm in a 1 second measurement, or to an SNR of 1.1 for the expected PV effect (~80 ppm) in a second of integration. The inset illustrates the relative size of the anticipated PV effect to the 1$\sigma$ error in the mean ratio over the 19 s of data acquisition.

With the SNR level currently obtained, a measurement of the PV-effect at the 1% level is expected to require a few hours of integration for a single isotope, where as a measurement at the 0.1% level will require substantially longer data acquisition. A further reduction of technical noise level in the apparatus, as well as an increase of the atomic population available to interact with the 408 nm light in the nonzero-isotopes (by optical pumping to a single ground-state Zeeman sublevel) will bring such a statistical accuracy within reach.

**5. Systematics**

The approach for handling systematic contributions to the PV measurements follows [2]. It starts with an analysis of the ratio of the 1st to 2nd harmonic amplitudes of the 408 nm rate in the presence of all possible field imperfections, including misalignments of the electric and magnetic fields with respect to the apparatus coordinate system, stray (non-reversing) components of these fields, as well as an ellipticity for the nominally linearly polarized light in the PBC. Several systematic terms then appear in the harmonics ratio, which are generally the product of two or more field imperfections. The majority of these terms do not transform like the Stark-PV signal under all field reversals combined (reversal of electric and magnetic field, light polarization rotation by $\pm\pi/2$), allowing us to discriminate the PV-signature against these. Only a few terms (just one in the $m=0 \rightarrow m´=0$ component of the 408 nm transition) mimic the PV-signal under all reversals, at the 0.1% level.

The scheme used to measure PV-mimicking systematics relies on enhancing one of the imperfections appearing in the systematic term and measuring the other. Although a detailed description of the routine employed in the Yb experiment will not be presented here, we do illustrate its essentials, taking as an example the $m=0 \rightarrow m´=0$ component of the 408 nm transition. An analysis for the ratio (10), combining measurements made under all possible configurations for the reversing fields, and keeping terms up to 2nd order in field imperfections, yields:

$$r = \frac{R^{1st}}{R^{2nd}} = \frac{4\zeta}{\beta E_0} + \frac{4 b_x^{ac} e_y}{B_z E_0}, \qquad (11)$$

where $b_x^{ac}$ is the reversing component of the leading field $B_z$, and $e_y$ is a stray electric field along the y-axis. The PV-mimicking term is expected to contribute by a few % to the ratio $r$ (given the degree of alignment of the magnetic-field coils with respect to the apparatus coordinate system, and a typical value of 1 V/cm for the stray field). A measurement of $b_x^{ac}$ can be done by imposing a large $e_y$ using the auxiliary field-plate electrodes, so that the 2$^{nd}$ term in (11) becomes sizable enough to be measured with good statistics. $b_x^{ac}$ can be then minimized using shimming-coils. With this method, it should be possible to reduce the contribution $b_x^{ac} e_y$ to a level lower than 0.1% of the PV-contribution. Conversely, enhancing $b_x^{ac}$ allows a measurement of $e_y$ and its potential drifts over time.

Several other exaggerated imperfections will also be imposed, to check the validity of the systematics model employed in the experiment. We do not anticipate that systematics will pose a limitation to the PV-sensitivity, down to the 0.1% level of accuracy.

**Conclusions**

We have presented progress towards making precise determinations of parity violation effects in Yb, with a newly constructed apparatus offering enhanced sensitivity. We have illustrated the interference technique we employ to carry out these measurements, described the experimental setup, and discussed the current sensitivity and our methods for studying systematics.

With our new apparatus, measurements of the PV-effect across a chain of isotopes will allow us to probe the variation of neutron distribution within the Yb nuclei and to probe for new couplings to protons and neutrons at the tree level, beyond the Standard Model. In addition, precise determination of the effect among the different hyperfine levels of the 408 nm transition in the nonzero-nuclear-spin isotopes, will probe their anapole, and help resolve long-lasting discrepancies in the parameter space of weak meson nucleon-nucleon coupling constants.

**Acknowledgements**

LB is supported by a Marie Curie Individual Fellowship within the second Horizon 2020 Work Programme. AF is a recipient of the Carl Zeiss Graduate Fellowship.